\documentclass{PoS}

\title{Determination of the analysing power for the $\vec{p}p \rightarrow pp\eta$ reaction using WASA-at-COSY detector system}

\ShortTitle{Study of the eta meson production with polarized proton}

\author{\speaker{I. Sch\"atti-Ozerianska}\thanks{for the WASA-at-COSY Collaboration.}\\
        Jagiellonian University\\
        E-mail: \email{i.ozerianska@gmail.com}}

\author{P.~Moskal\\
        Jagiellonian University\\
       E-mail: \email{p.moskal@uj.edu.pl}}
        
\author{ M. Zieli\'{n}ski\\
        Jagiellonian University\\
       E-mail: \email{marcin.zielinski@uj.edu.pl}}
        
\abstract{We report on the measurement of the analyzing power for the $\vec{p}p\to pp\eta$ reaction with beam momenta of 2026~MeV/$c$ and 2188~MeV/$c$ performed with the WASA-at-COSY detector at the Cooler Synchrotron COSY.  The $\eta$ meson from the $\vec{p}p\to pp\eta$ reaction was identified by the techniques of missing mass and invariant mass.
The angular distribution of the determined analyzing power strongly disagrees
 with theoretical predictions. A comparison of the obtained $A_y$ angular distribution with a series of associated Legendre polynomials revealed negligible contribution of the $Sd$ partial wave at Q~=~15~MeV. However, at Q~=~72~MeV, a significant interference of the $Ps$ and $Pp$ partial waves was observed.}

\FullConference{54th International Winter Meeting on Nuclear Physics\\
		25-29 January 2016 \\
		Bormio, Italy}

\begin{document}

\section{Introduction}
The production mechanism of the $\eta$ meson and meson-nucleon final state interaction for the $\vec{p}p\to pp\eta$ reaction
can be studied via measurements of the cross sections and analyzing power, $A_{y}(\theta)$. Up to now total and differential cross sections have been determined relatively precisely~\cite{chiavassa94,calen96,calen97,hibou98,smyrski00,bergdolt93,abdelbary03,
ETA-PRC-Moskal,ETA-EPJ-Moskal,ETA-Petren,pn-pneta-Calen,pn-pneta-Moskal}, however so far $A_{y}$ for the $\vec{p}p\to pp\eta$ reaction has been determined with rather large uncertainties~\cite{rafalprl,disto,pwinter,pwinter1}. 

In November 2010 the high statistics sample of $\vec{p}p\to pp\eta$ reaction has
 been collected using the azimuthally symmetric WASA-at-COSY detector \cite{spin2010}. Measurements were taken with two beam momenta of 2026 MeV/$c$ and 2188 MeV/$c$, corresponding to 15 MeV and 72 MeV excess energies, respectively.
 
 Based on elastic scattering of protons, the vertex position of the real experiment were measured with two independent methods \cite{malg}. The spin flipping technique of the beam has been used to control the effect caused by potential asymmetries in the detector.
 Monitoring of the beam polarization was based on the $\vec{p}p\to pp$ reaction. The result shown stable polarization during whole experiment \cite{pol,pol1}.
 
\section{Analyzing power for the $\eta$ meson}
The determination of the analyzing power for the $\eta$ meson was carried out separately for spin up and spin down modes, and for each spin orientation the analyzing power was determined identifying the $\eta$ meson via two decay channels: $\eta\rightarrow \gamma\gamma$ and  $\eta\rightarrow 3\pi^0$.

\begin{figure}[h!]
    \label{fig:gea}
    \includegraphics[width =0.5\textwidth, height=5.3cm]{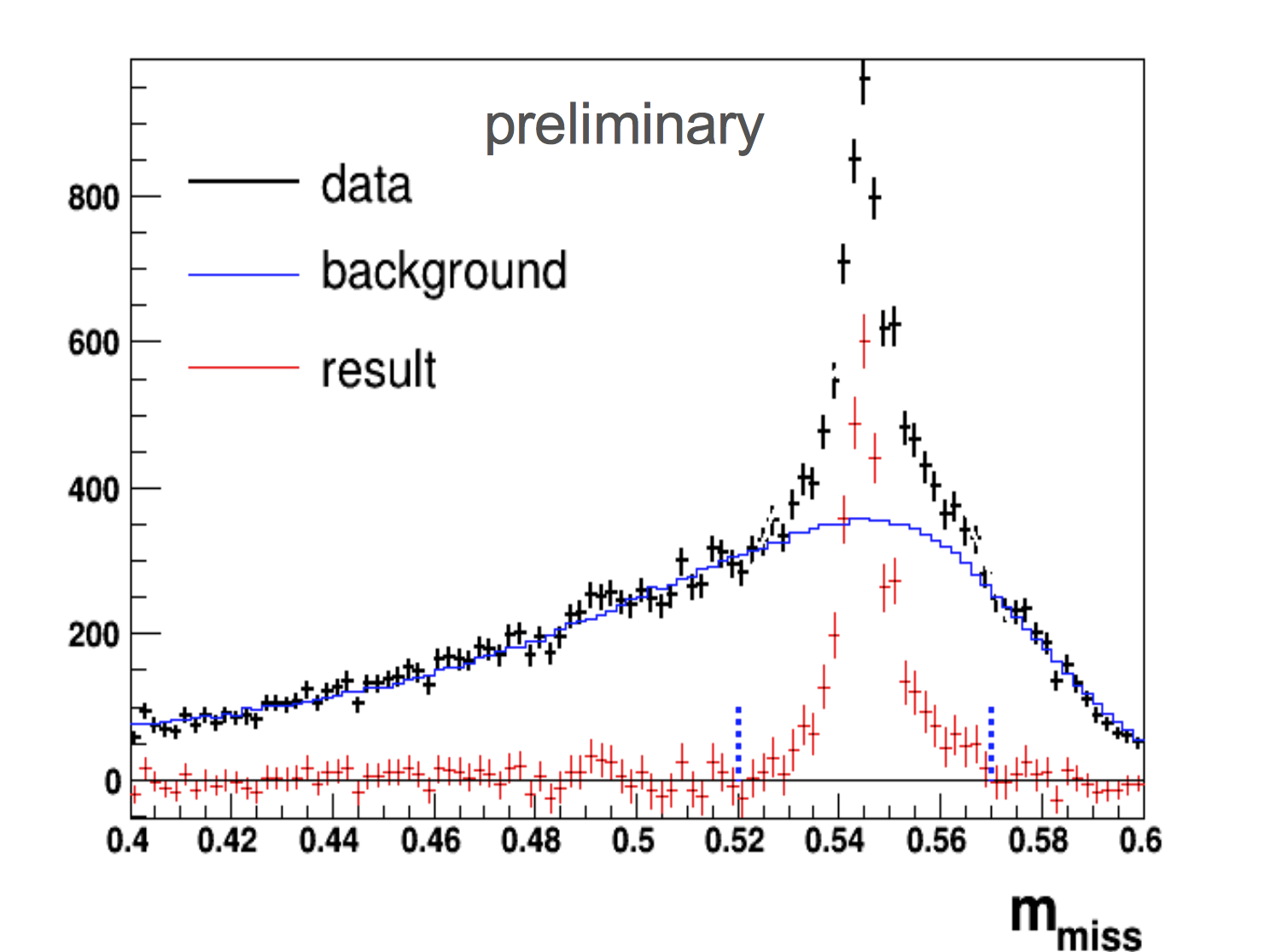}
\quad
    \label{fig:gea}
    \includegraphics[width =0.5\textwidth, height=5.3cm]{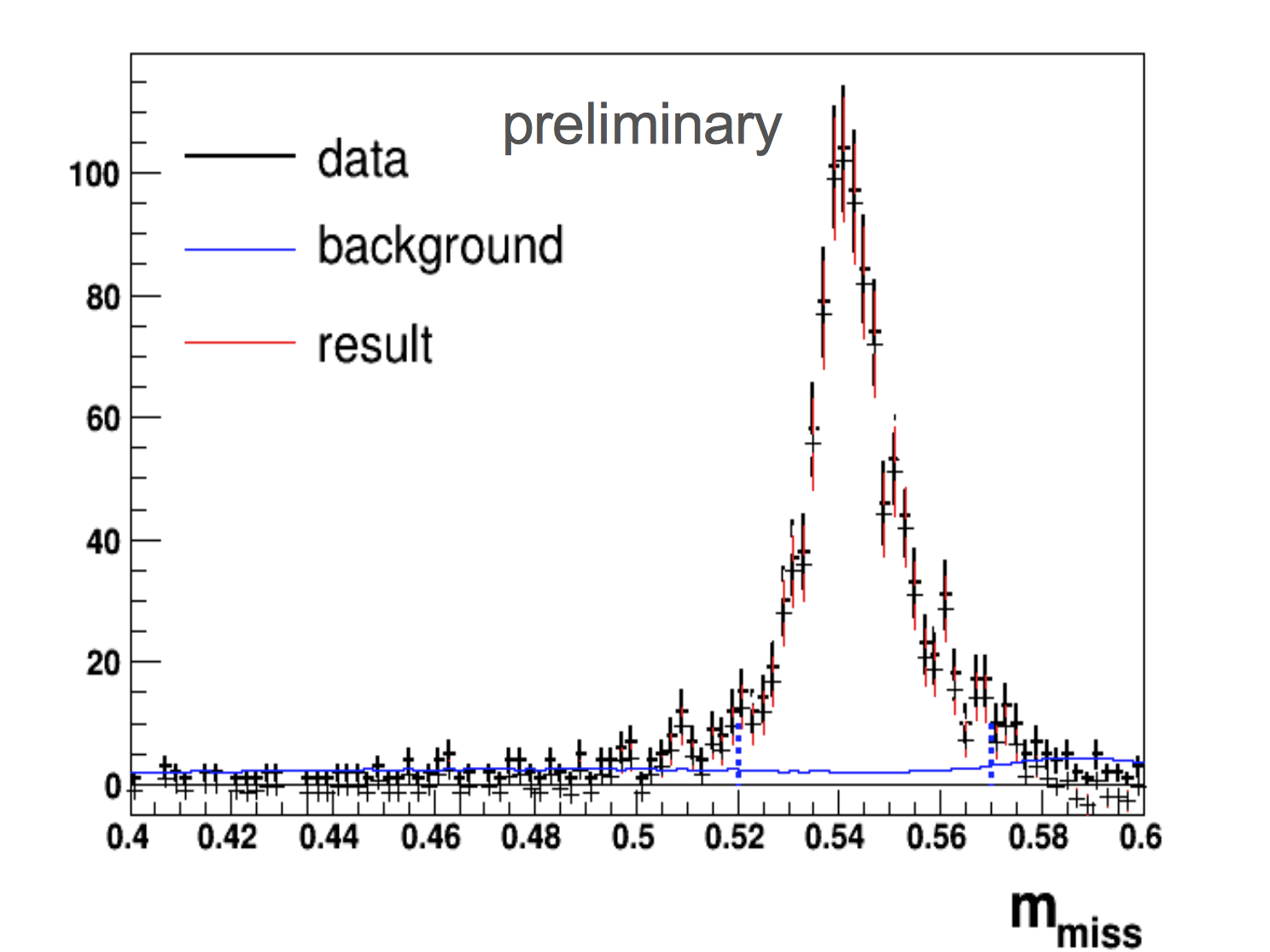}

 \caption{ Missing mass distribution for the chosen range $70^{\circ}<\theta_{\eta}<90^{\circ}$, $-180^{\circ}<\varphi_{\eta}<-170^{\circ}$ and spin "up" mode. Left: $\eta \rightarrow \gamma \gamma$. Right: $\eta \rightarrow 3\pi^0 \rightarrow 6 \gamma$. Beam momentum: $p_{beam} = 2188$~MeV/$c$. Black crosses denote experimental data. Continuous blue lines show the sum of the simulated background for the $\pi^0,2\pi^0,3\pi^0$ and $4\pi^0$ production. Red points show the result of difference between the experimental data and simulated background. Dashed blue lines show the region of the extraction of the number of produced $\eta$ meson.}

  \label{ex_2g_2188}
\end{figure}

After the identification of the final state particles the number of events corresponding to the \\$\vec{p}p\to pp\eta$ reaction, 
have been determined for each angular bin N($\theta_{\eta}$,$\varphi_{\eta}$) separately. $\theta_\eta$ and $\varphi_\eta$ denote respectively the polar and the azimuthal angle of the eta meson emission in the center  of mass system. 
An example of the missing mass distribution for a chosen spin mode of the beam momentum 2188~MeV/$c$ is shown in Figure~\ref{ex_2g_2188}.


\begin{figure}[h!]
    \label{fig:gea}
    \includegraphics[width =0.5\textwidth, height=5.3cm]{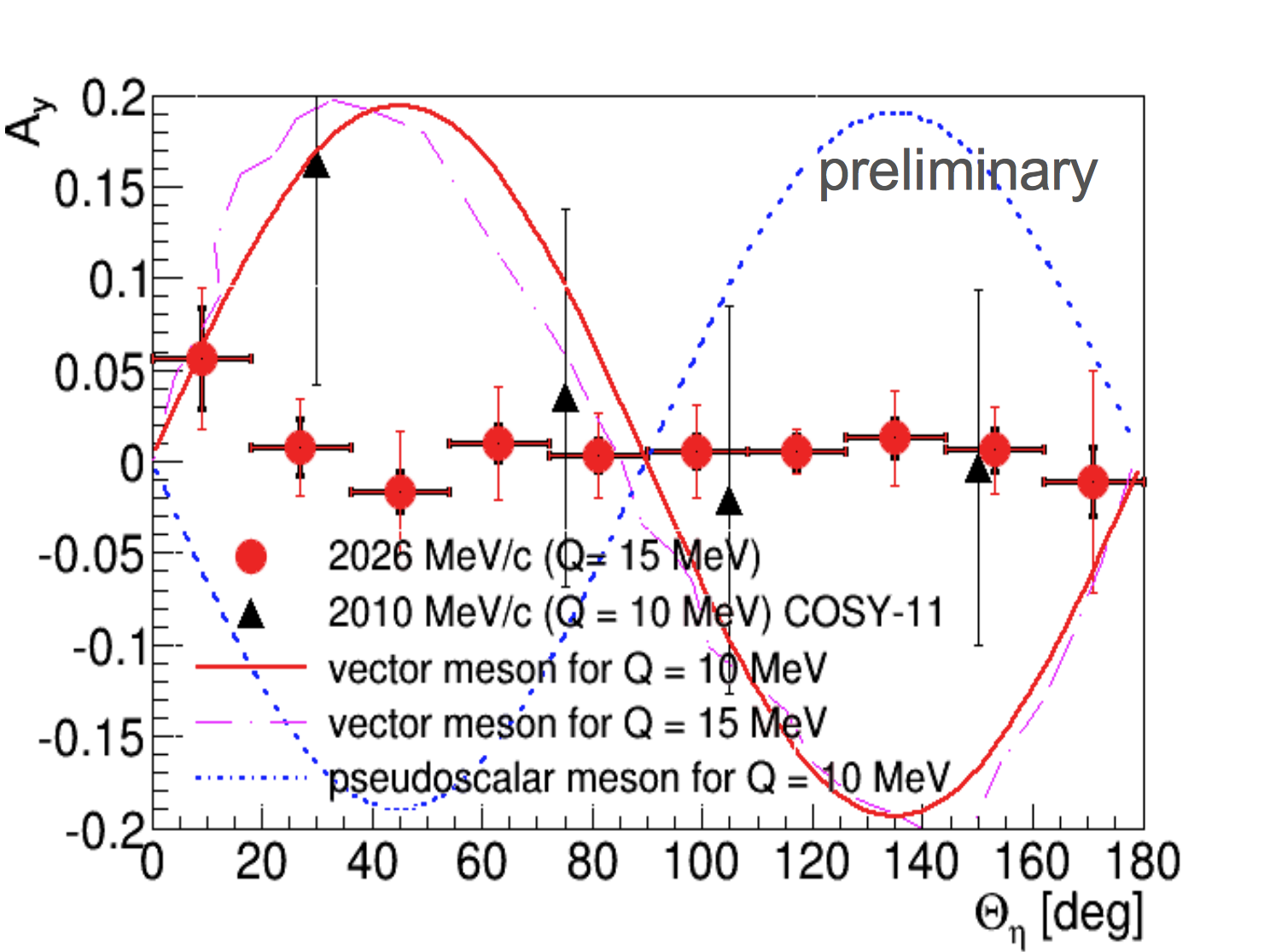}
\quad
    \label{fig:gea}
    \includegraphics[width =0.5\textwidth, height=5.3cm]{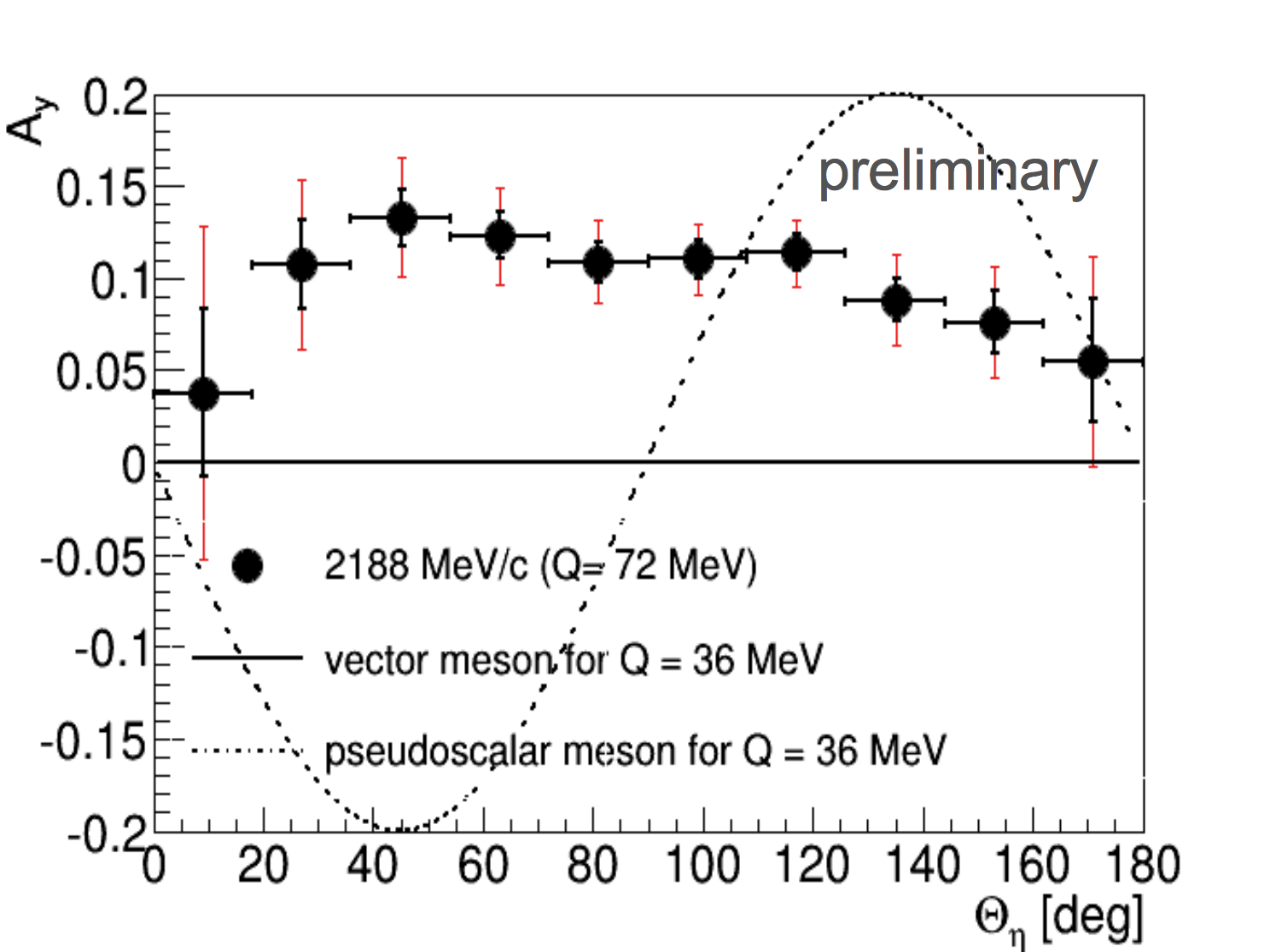}

 \caption{ Analyzing power of the $\eta$ meson as a function of $\theta_{\eta}$. Superimposed lines indicate theoretical predictions (see legend) for Q = 15 MeV (left panel) and Q = 72 MeV (right panel).The dashed line shows the prediction of the analyzing power as a function of the $\eta$ emission angle in the center-of-mass frame for the vector meson dominance model \cite{nakayama03}. The solid line describes the vector meson 
model~\cite{faldt01} and the dotted line describes the pseudoscalar model~\cite{nakayama02}. Please note that the data at Q = 72~MeV are compared with theoretical predictions for Q = 36~MeV, which is the largest Q for which such predictions are available. }

  \label{theta_result_2_explanation}
\end{figure}
The collected amount of the $\eta$ events, about 400 000 events, significantly improves the statistical uncertainty of the analyzing power for the $\eta$ meson compared to the previous COSY-11 experiments with about 2000 events only ~\cite{rafalprl}. The systematic uncertainty was improved due to the axial symmetry of the WASA-at-COSY detector and its close to 4$\pi$ acceptance which is by two orders of magnitude larger than the acceptance of the COSY-11 detector.

\begin{figure}[h!]
    \label{fig:gea}
    \includegraphics[width =0.5\textwidth, height=5.3cm]{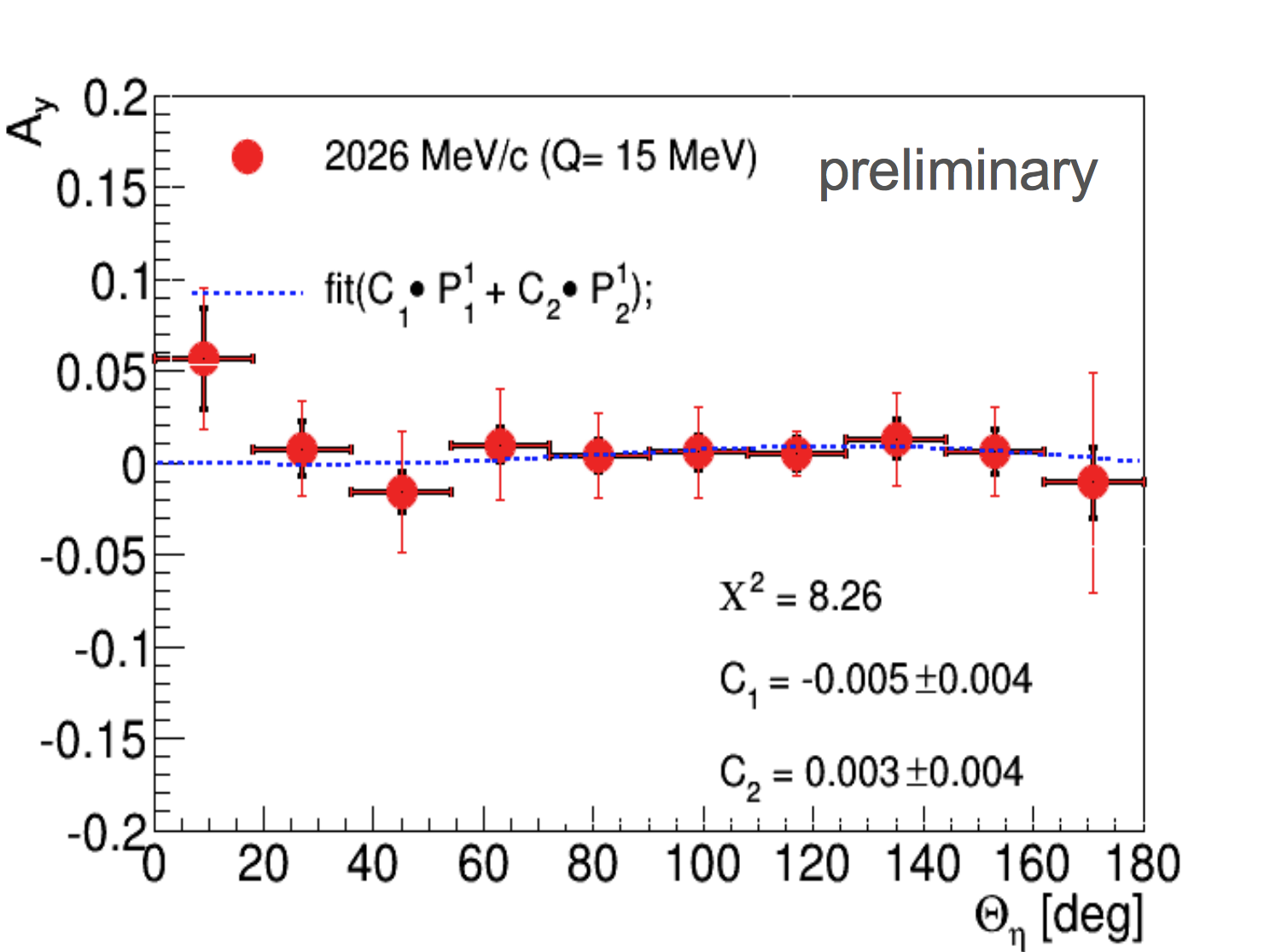}
\quad
    \label{fig:gea}
    \includegraphics[width =0.5\textwidth, height=5.3cm]{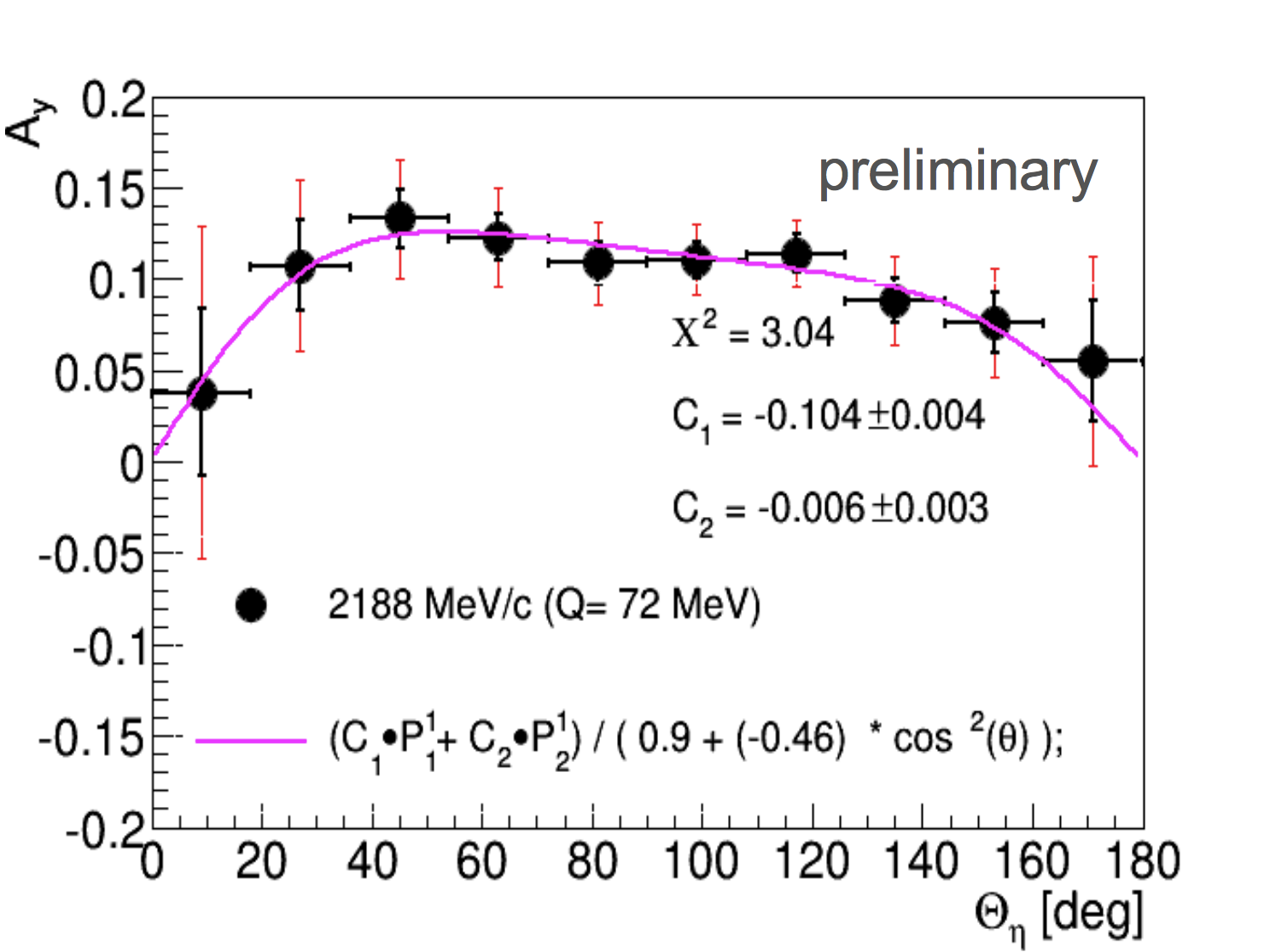}

 \caption{ Analyzing power of the $\eta$ meson as a function of $\theta_{\eta}$. The fit of $A_y$ with the sum of the two associated Legendre polynomials $P^1_1$ and $P^1_2$ is shown for the Q~=~15 MeV (left) and for Q~=~72 MeV (right).}

  \label{theta_result_1leg}
\end{figure}

Assuming that $p$ and $d$ waves can occur for the $\eta$ meson production, its analyzing power is given by:
\begin{equation}
A_y = \frac{\Im (A_{Ps}A^*_{Pp})sin\theta_{\eta} + \Im(A_{Ss}A^*_{Sd})3cos\theta_{\eta}\sin\theta_{\eta}}{\frac{d\sigma}{d\Omega}},
\label{Ay_pwd}
\end{equation}
where $\Im (A_{Ps}A^*_{Pp})$ is the imaginary part of the interference term between the $Ps$ and $Pp$ waves, and $\Im(A_{Ss}A^*_{Sd})$ is the interference term between the $Ss$ and $Sd$ waves \cite{saha}.
Our experimental results are shown in Figure~2 
together with meson exchange model predictions.
Figure~\ref{theta_result_1leg} shows result obtained in this experiment with superimposed lines corresponding to the fit of the formula:
 
 \begin{equation}
 A_y\frac{d\sigma}{d\Omega}=C_1\cdot sin\theta_{\eta}+C_2\cdot cos\theta_{\eta} sin\theta_{\eta}
 \end{equation}
   where $C_1$ and $C_2$ are treated as free parameters of the fit.
   For Q~=~72~MeV the angular dependence of $d\sigma/d\Omega$ was determined by the parametrization of the data from reference \cite{ETA-Petren}, 
and for Q~=~15~MeV it was assumed to be constant as determined in the experiments of COSY~11~\cite{ETA-PRC-Moskal}
and COSY-TOF collaborations~\cite{abdelbary03}.   
One can see in Figure~\ref{theta_result_1leg} that the associated Legendre polynomials of order $m = 1$ fully describe the existing data. 

Thus, the analyzing power is zero for the beam momentum 2026~MeV/$c$, and there is no interference between $A_{Ss}$ and $A_{Sd}$ as well as between $A_{Pp}$ and $A_{Ps}$ amplitudes of the partial waves.

\section{Results}

The comparison of the angular dependence of the analyzing power for the $\vec{p}p\to pp \eta$ reactions with the associated Legendre polynomials revealed that at Q=15 MeV there is no $Ss-Sd$ and no $Pp-Ps$ interference and that for the higher beam momentum 2188~MeV/$c$, the $Sd$ partial wave contribution is small (consistent with zero within two standard deviations). On the other hand, the contribution of $Ps-Pp$ interference is large which means that both of these partial waves contribute at Q~=~72~MeV (see Figure~\ref{theta_result_1leg}).

The obtained angular dependence of the analyzing power agrees with the previous experiments, however it disagrees with the theoretical predictions based on the pseudoscalar or vector meson dominance models~\cite{faldt01,nakayama02}.




\vspace{0.3cm}
\section{Acknowledgment}
We acknowledge support by the Polish National Science Center through
grant No.~2011/03/B/ ST2/01847, by the FFE grants of the Research Center
 Juelich, by the EU Integrated Infrastructure Initiative HadronPhysics
Project under contract number RII3-CT-2004-506078 and by the European
 Commission under the 7th Framework Programme through the Research Infrastructures action of the Capacities Programme, Call: FP7-
INFRASTRUCTURES-2008-1, Grant Agreement N. 227431.


\vspace{-0.1cm}

\medskip $^a$ M.~Smoluchowski Institute of Physics, Jagiellonian University,
30-059 Cracow, Poland\\

\end{document}